\documentclass[reprint,english,aps,prl,showpacs,superscriptaddress,preprintnumbers]{revtex4-1}
\usepackage[T1]{fontenc}
\usepackage[latin9]{inputenc}
\usepackage{float}
\usepackage{amsmath}
\usepackage{amssymb}
\usepackage{graphicx}
\usepackage{esint}
\usepackage{natbib}
\usepackage{xcolor}

\makeatletter
\@ifundefined{textcolor}{}
{%
 \definecolor{BLACK}{gray}{0}
 \definecolor{WHITE}{gray}{1}
 \definecolor{RED}{rgb}{1,0,0}
 \definecolor{GREEN}{rgb}{0,1,0}
 \definecolor{BLUE}{rgb}{0,0,1}
 \definecolor{CYAN}{cmyk}{1,0,0,0}
 \definecolor{MAGENTA}{cmyk}{0,1,0,0}
 \definecolor{YELLOW}{cmyk}{0,0,1,0}
}

\usepackage{amstext}
\usepackage{babel}
\newcommand{\unit}{1\!\!1}
\newcommand{\comment}[1]{}
\addto\captionsenglish{}
\makeatother

\begin{document}
\title{Entangled light from bimodal optical nanoantennas}
\author{J. Straubel}
\affiliation{Institute of Theoretical Solid State Physics, Karlsruhe Institute of Technology, 76131 Karlsruhe, Germany}
\author{R. Sarniak}
\affiliation{Centre for Astronomy, Faculty of Physics, Astronomy and Informatics, Nicolaus Copernicus University, Grudziadzka 5, 87-100 Torun, Poland}
\author{C. Rockstuhl}
\affiliation{Institute of Theoretical Solid State Physics, Karlsruhe Institute of Technology, 76131 Karlsruhe, Germany}
\affiliation{Institute of Nanotechnology, Karlsruhe Institute of Technology, 76131 Karlsruhe, Germany}
\author{K. S\l owik}
\email{karolina@fizyka.umk.pl}
\affiliation{Institute of Physics, Faculty of Physics, Astronomy and Informatics, Nicolaus Copernicus University, Grudziadzka 5, 87-100 Torun, Poland}

\begin{abstract}
We suggest a hybrid plasmonic device made of a bimodal metallic nanoantenna coupled to an incoherently pumped quantum emitter. This device emits light into the two modes entangled in the number of photons. The process is a prime example for losses turning from a nuisance into something beneficial, since, even though counterintuitively, the entanglement is enabled by strong incoherent processes, i.e. dominant scattering and absorption rates of the nanoantenna. This renders the nanoantenna an active source of nonclassicality. Both, the high emission rate and the degree of entanglement of the emitted light are insensitive with respect to imperfections in the nanoantenna length, rendering the scheme feasible for an implementation.
\end{abstract}


\maketitle
\section{Introduction} \label{sec:introduction}

Plasmonic and dielectric nanoantennas provide a platform for controlling light-matter interactions at the nanoscale due to their ability to tailor spatial and spectral properties of electromagnetic fields with high precision \cite{Muehlschlegel2005,Alu2008,Luk2010,Staude2013}. Additionally, quantum emitters can be accurately positioned at electromagnetic hot-spots localized at the nanoantenna vicinity \cite{Bleuse2011,Lehr2014}.\comment{,Chekini2015.} The resulting huge values of coupling between the quantum emitters and the nanoantenna via electromagnetic fields can be exploited, e.g. to boost molecular fluorescence \comment{Anger2006,} \cite{Taminiau2008,Krajnik2013,Mohtashami2015}, to enhance dipole-dipole interactions between multiple emitters \cite{Dzsotjan2010,Hou2014}, or to create entanglement between them \cite{Gonzalez2011,Lee2013}. The properties of radiated light can be modified as well, e.g. due to surface-enhanced Raman spectroscopy \cite{Fleischmann1974,Schmidt2016}. Devices made of nanoantennas coupled to appropriate quantum emitters have been proposed and exploited as sources of single photons \cite{Schietinger2009,Esteban2010,Filter2014}, or even pairs of photons entangled in the polarization degree of freedom \cite{Maksymov2012}.

Please note that in the latter and other contributions that discuss nonclassical light emission in nanoplasmonic cavities, the role of the nanoantennas is rather passive: It is the quantum emitter that provides the nonclassical statistics or entanglement, and the nanoantenna is merely used to boost the speed of the discussed processes. On the contrary, in the scheme proposed here nanoantennas are utilized as active sources. They are at the origin of the nonclassicality. The entanglement of the emitted light is a result of the bimodal character of the nanoantenna emission spectrum. Without the nanoantenna, no entanglement would be present at all.

\begin{figure}[h!]
\begin{centering}
\includegraphics[width=8.0cm,keepaspectratio]{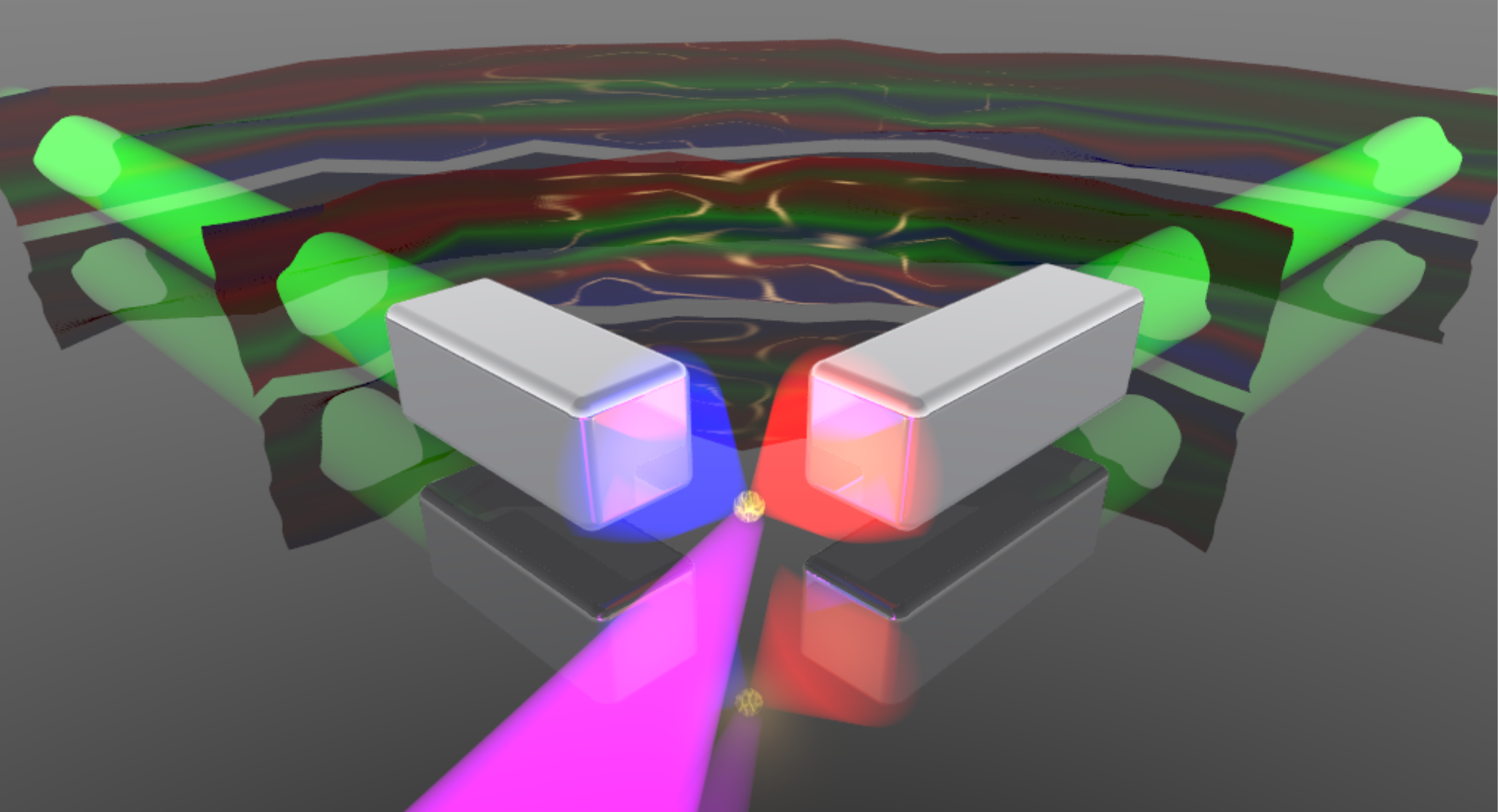}
\par
\end{centering}
\caption{\label{fig:illustration} Illustration of the generation of entangled light based on the bimodal properties of a nanoantenna.}\end{figure}

We discuss metal nanoantennas with spectra characterized by two overlapping resonances, coupled to an adjacent quantum emitter. Such a hybrid plasmonic device emits light in two modes that are entangled in the occupation number [Fig.~\ref{fig:illustration}]. Depending on the shape of the nanoantenna, the emission of the two modes might differ in the geometry of their emission patterns, the polarization, or the involved emission frequencies. We emphasize that similar bimodal nanoantennas have recently been fabricated \cite{dopf2015,black2014}. We base our work on prior contributions \cite{Plenio2002}, in which a similar scheme based on lossy bimodal cavities coupled to a two-level atomic system was proposed. There, significant noise of the cavity was crucial to obtain entanglement of the emitted light. The huge coupling of nanoantennas and quantum emitters is realized at even larger photon scattering and absorption rates. This promises bright entangled light, i.e. of GHz emission rates. The emitted states may be used for quantum computing with spatially delocalized qubits \cite{Mompart2003}, as Bell states for quantum information processing \cite{Hardy1994,Hessmo2004}, or as a basis to prepare multipartite entanglement \cite{Mompart2003}.

Please note that the quantum emitter is the source of photons in our setup, capable of emitting only one photon at a time. Due to the presence of the nanoantenna of a bimodal character, these photons are emitted with very high probability in the two nanoantenna modes. This means each of the modes contains either one or no photons. Even though we do not know which of the modes is occupied, we know that if one of them is, the other must not be. This is a special kind of correlation, for which in a sense more is known about the total system than of its individual parts. This in itself is not yet nonclassical: the same kind of correlations characterize a pair of shoes. The nonclassical spirit comes from the fact that the state of each of these photons is a coherent superposition of the two possibilities, and can be classified as "mode 1" or "mode 2" photon only after a measurement is performed and the state collapses. Such quantum correlations are called entanglement. It should be emphasized that such considerations can only be made in presence of the nanoantenna, which defines the emission modes, and provides means to perform an experiment, i.e. due to its directive properties. For this reason, we call the nanoantenna the source of nonclassicality for photons originating from the quantum emitter. 

The document is structured as follows. In Section \ref{sec:principle}, the operating principle of the source is shortly introduced. Next, an experimentally feasible set of bimodal nanoantennas is proposed in Section \ref{sec:nanoantenna}, with the corresponding scattering and absorption characteristics due to a dipolar excitation that represents a quantum emitter. The photonic emission rate and the degree of entanglement of light emitted by the hybrid system of a bimodal nanoantenna and a quantum emitter are calculated in Section \ref{sec:quantum}, and expressed in function of experimentally relevant parameters. Three appendices follow: The method that we have used to obtain the nanoantenna spectra is explained in Appendix A. The following Appendix B includes an introduction of the quantities characterizing the nanoantenna's optical properties, in terms of parameters that enter the quantum Lindblad equation, which provides the state of the emitted light. In Appendix C, details related to the quantum calculation are provided.


\section{Operating principle} \label{sec:principle}

\begin{figure}[h!]
\begin{centering}
\includegraphics[width=8.6cm,keepaspectratio]{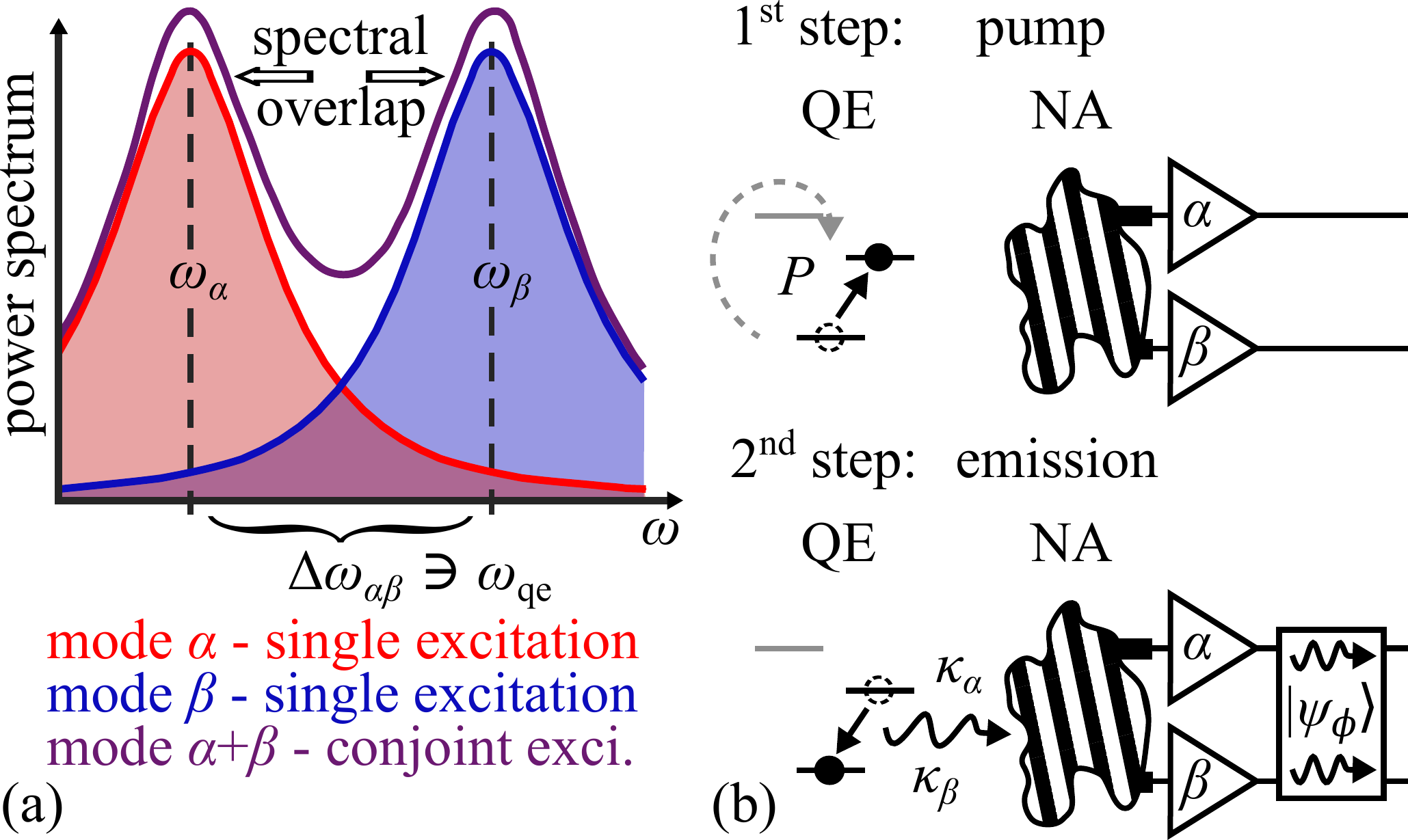}
\par
\end{centering}
\caption{\label{fig:nanoantenna1} (a) Generic spectrum of a considered doubly resonant nanoantenna that features two modes at resonance frequency $\omega_{\alpha}$ and $\omega_{\beta}$. Their spectral overlap $\Delta\omega_{\alpha\beta}$ contains a quantum emitter with a transition frequency of $\omega_{\mathrm{qe}}$.
(b) Schematic depicting the two step operation principle of the hybrid system consisting of nanoantenna (NA) and quantum emitter (QE). Included are pump $P$, bimodal coupling with coupling constants $\kappa_{\alpha}$ and $\kappa_{\beta}$, and entangled emission of $|\psi_\phi\rangle$ in modes $\alpha$ and $\beta$. To keep the emission channel well separated from the pump beam, a third auxiliary level of the quantum emitter is used to realize the pumping step.}\end{figure}

The entangled light source consists of a coupled system made from a quantum emitter with two active levels and a doubly resonant nanoantenna. Each resonance is described by a quantum-mechanical bosonic mode. Since the quantum emitter shall couple to both modes, we require its transition frequency to be contained inside the spectral overlap of the modes [Fig.~\ref{fig:nanoantenna1}(a)]. In the first step of the light generation procedure, the quantum emitter is pumped incoherently into its excited state [Fig.~\ref{fig:nanoantenna1}(b)]. Next, it may emit a photon into each of the nanoantenna modes with probabilities related to the corresponding coupling constants. This produces an entangled state of light of the form $|\psi_\phi\rangle = \left(|10\rangle + e^{i\phi}|01\rangle\right)/\sqrt{2}$. The numbers inside the kets denote photon numbers in each of the modes. The design of such a device and its performance are the main aspects described in the following.

To meet the requirements of the scheme, we design nanoantennas whose scattering/absorption spectra can be well characterized by two overlapping but uncoupled resonances. Based on the spectra, the scattering/absorption rates and the coupling strengths to given quantum emitters in their vicinity can be calculated. This information is fuelled into equations of motions for the quantum mechanical system. This allows to obtain the stationary density matrix of the system and the corresponding degree of entanglement between the photonic modes.

\section{Bimodal nanoantenna}\label{sec:nanoantenna}
\begin{figure}[h!]
\begin{centering}
\includegraphics[width=8.6cm,keepaspectratio]{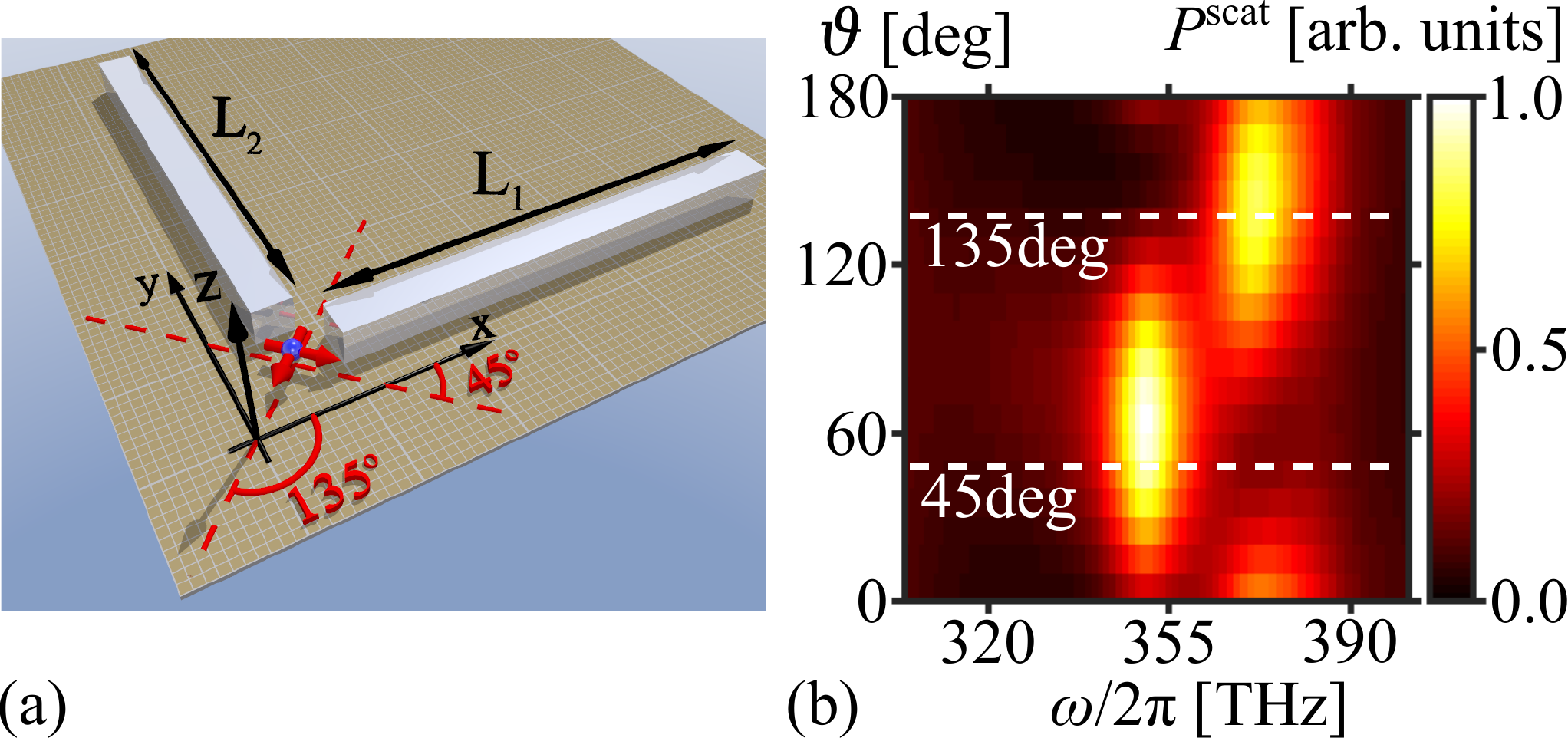}
\par
\end{centering}
\caption{\label{fig:nanoantenna2} (a) Schematic of the nanoantenna. The blue dot corresponds to the position of the quantum emitter.
(b) Scattering spectra $P^\mathrm{scat}(\omega)$ for quantum emitter orientations $\vartheta$ in $\left[0^{\circ}, 180^{\circ}\right]$ for $\Delta L=90$ nm. The two orientations corresponding to the single excitation are marked in white.}
\end{figure}

The proposed nanoantenna that copes with all these requirements consists of two cuboidal silver nanorods, of lengths $L_1 = 250$nm, and $L_2 = L_1-\Delta L$, where $\Delta L \in [0,L_1/2]$. The nanorods have square cross sections of 20nm $\times$ 20nm. They are embedded in glass with a relative permittivity $\varepsilon=2.25$ and their long axes are oriented perpendicular to each other, forming a L-shaped structure [Fig.~\ref{fig:nanoantenna2}(a)].

Such a nanoantenna design requires the precise, equidistant placement of the quantum emitter with respect to the two nanorods as indicated in Fig.~\ref{fig:nanoantenna2}(a). But with a quantum emitter located according to the requirements, the coupling to the two different modes can be controlled by the orientation of the transition dipole moment of the quantum emitter.

We first analyze the scattering spectra of such a nanoantenna due to a dipole emitter placed in the vertex of the symmetry axes of the two nanorods [Fig.~\ref{fig:nanoantenna2}(b)]. The spectra have been obtained for $\Delta L=90$nm and for several orientations of the dipole, between $\vartheta = 0^{\circ}$ and $\vartheta = 180^{\circ}$ with respect to the $x$-axis of the reference frame in Fig.~\ref{fig:nanoantenna2}(a). Our calculation method is described in detail in Appendix A. \comment{Ref.~\onlinecite Straubel2016} Such scans reveal two uncoupled but overlapping resonances at optical frequencies that fulfil the requirements imposed on the nanoantenna. The two modes can be addressed individually by a dipole oriented in one of the two basic directions, corresponding to the rotation angles of $\vartheta = 45^{\circ}$ and $\vartheta = 135^{\circ}$. For any other orientation, both resonances are excited simultaneously.

\begin{figure}[h!]
\begin{centering}
\includegraphics[width=8.6cm,keepaspectratio]{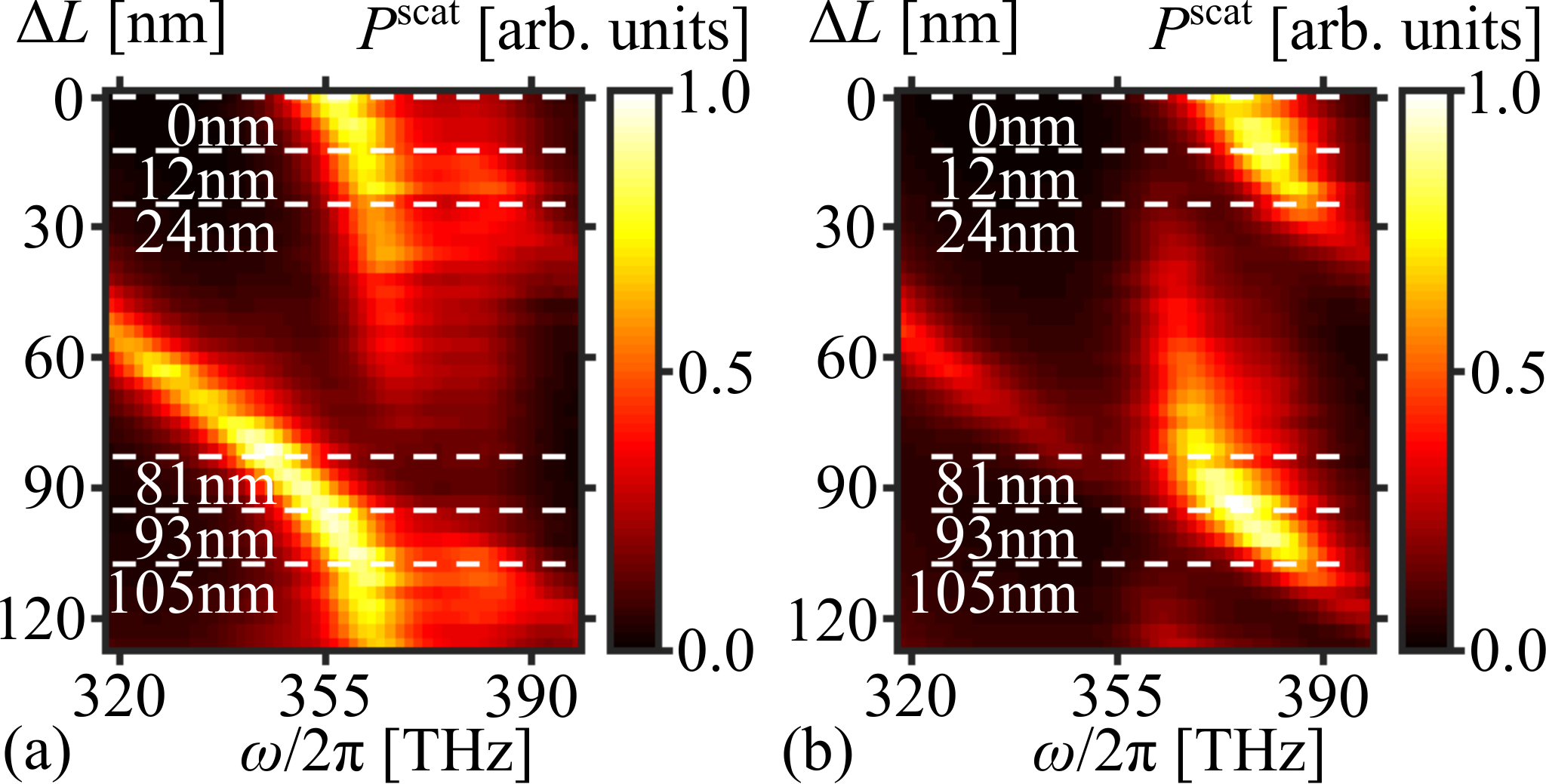}
\par
\end{centering}
\caption{\label{fig:nanoantenna3} (a) Scattering spectra $P^\mathrm{scat}(\omega)$ for quantum emitters with $\vartheta=45^{\circ}$ and varying $\Delta L$. 
(b) As in (a), but with $\vartheta=135^{\circ}$. White dashed lines in (a) \& (b) mark the lengths differences selected for further analysis.}
\end{figure}

We continue to investigate the scattering properties of the nanoantenna for the two basic dipole orientations, corresponding to the individual mode excitation with $\Delta L = 90$nm [Fig.~\ref{fig:nanoantenna3}(a)\&(b)], by scanning the lengths difference $\Delta L$ of the two branches from equally long to a length ratio of $1:2$, while the equidistant location of the quantum emitter with respect to the two nanorods is preserved. As a result, we find two pairs of resonances present for two distinct ranges of the lengths difference: for $\Delta L\in [0,30\mbox{nm})$ and for $\Delta L\in (70\mbox{nm},120\mbox{nm})$. This indicates that the desired behavior is stable with respect to the nanoantenna length, being beneficial for a possible fabrication. In the scattering [Fig.~\ref{fig:nanoantenna3}(a)\&(b)] and absorption [Fig.~\ref{fig:AppA}(a)\&(b) as shown in Appendix A] spectra we observe a pronounced anticrossing. Furthermore, from the requirement of having a dipole orientation of $\vartheta=45^{\circ}$ and $\vartheta=135^{\circ}$ to have decoupled resonances at two different frequencies, we conclude that the two modes we capitalize on emerge from the hybridization of modes sustained by the individual nanorods. Hybridization occurs between modes of either the same or adjacent orders, depending on the chosen length regime.

For the further analysis, we have chosen six selected lengths differences $\Delta L$: three in each of the two ranges. The selected values of $\Delta L$ read: $0$nm, $12$nm, $24$nm, and $81$nm, $93$nm, $105$nm. They are marked in Fig.~\ref{fig:nanoantenna3}(a)\&(b) with white dashed lines.

In the next step, we fit the scattering and absorption spectra of the selected nanoantennas with a pair of Lorentzian resonances for each of the selected $\Delta L$s, 
according to the procedure described in detail in Appendix B. \comment{Ref.~\onlinecite Straubel2016} Based on such fits, we obtain the central frequencies $\omega_\mathrm{j}$ for each mode $\mathrm{j}=1,2$, \comment{. The full-width at half-maximum of each fitted Lorentzian indicates,} the far-field scattering rate $\Gamma^\mathrm{scat}_\mathrm{j}$ and the bulk absorption rate $\Gamma^\mathrm{abs}_\mathrm{j}$ for the given mode $\mathrm{j}$. The corresponding mode efficiency is defined as $\eta_\mathrm{j} =\Gamma^\mathrm{scat}_\mathrm{j}/\Gamma_\mathrm{j}$, where $\Gamma_\mathrm{j}=\Gamma^\mathrm{scat}_\mathrm{j}+\Gamma^\mathrm{abs}_\mathrm{j}$. The Purcell factor $F$ indicates the ratio of the power emitted by the dipole source supported by the $\mathrm{j}^\mathrm{th}$ nanoantenna mode, and the power emitted in the absence of the nanoantenna \cite{Devilez2010,Straubel2016}. A comparison of Purcell factors obtained in this way with the corresponding cavity-QED expressions \cite{Koppens2011,Straubel2016}, leads to an estimation of the coupling constants $\kappa_\mathrm{j}$ of the nanoantenna modes to the quantum emitter (Appendix B). This provides a set of parameters characterizing the modes of the nanoantenna, which should be treated quantum-mechanically in the single-photon regime. 
These parameters are gathered in Table~\ref{tab:parameters}. They enter the quantum-mechanical equations in the next Section.

\begin{table}[h]
{
\begin{tabular*}{8.6cm}{@{\extracolsep{\fill}}|c|c|c|c|c|c|c|}
\hline
\begin{tabular}{c}
$\Delta L$\\
$[\mathrm{nm}]$
\end{tabular}&
\begin{tabular}{c} 
mode\\
no. $\mathrm{j}$
\end{tabular}& 
\begin{tabular}{c}
$\omega_\mathrm{j}/2\pi$ \\
$[\mathrm{THz}]$
\end{tabular}&
\begin{tabular}{c}
$\Gamma^\mathrm{scat}_\mathrm{j}/2\pi [\mathrm{THz}]$,\\
$\Gamma^\mathrm{abs}_\mathrm{j}/2\pi [\mathrm{THz}]$
\end{tabular}&
$\eta_\mathrm{j}$ [$\%$]
&
\begin{tabular}{c}
$\kappa_\mathrm{j}/2\pi$\\
$[\mathrm{THz}]$
\end{tabular}&
$F$\\
\hline
0 & 
\begin{tabular}{c}
$1$\\
$2$
\end{tabular}&
\begin{tabular}{c}
$358$\\ 
$374$
\end{tabular}& 
\begin{tabular}{c}
$15.0$, $17.7$\\
$16.1$, $18.8$
\end{tabular}& 
\begin{tabular}{c}
$45.9$\\
$46.1$
\end{tabular}& 
\begin{tabular}{c}
$0.18$\\
$0.13$
\end{tabular}& 
\begin{tabular}{c}
$5.73$\\ 
$5.27$
\end{tabular}\\
\hline

12 & 
\begin{tabular}{c}
$1$\\
$2$
\end{tabular}&
\begin{tabular}{c}
$361$\\ 
$379$
\end{tabular}& 
\begin{tabular}{c}
$14.5$, $17.2$\\
$13.7$, $19.1$
\end{tabular}& 
\begin{tabular}{c}
$45.6$\\
$41.7$
\end{tabular}& 
\begin{tabular}{c}
$0.17$\\
$0.10$
\end{tabular}& 
\begin{tabular}{c}
$5.63$\\ 
$4.87$
\end{tabular}\\
\hline

24 & 
\begin{tabular}{c}
$1$\\
$2$
\end{tabular}&
\begin{tabular}{c}
$364$\\ 
$387$
\end{tabular}& 
\begin{tabular}{c}
$17.0$, $17.3$\\
$15.0$, $22.1$
\end{tabular}& 
\begin{tabular}{c}
$49.5$\\
$40.3$
\end{tabular}& 
\begin{tabular}{c}
$0.15$\\
$0.07$
\end{tabular}& 
\begin{tabular}{c}
$5.28$\\ 
$4.49$
\end{tabular}\\
\hline

81 & 
\begin{tabular}{c}
$1$\\
$2$
\end{tabular}&
\begin{tabular}{c}
$345$\\ 
$369$
\end{tabular}& 
\begin{tabular}{c}
$15.0$, $16.6$\\
$16.1$, $16.7$
\end{tabular}& 
\begin{tabular}{c}
$47.5$\\
$49.0$
\end{tabular}& 
\begin{tabular}{c}
$0.12$\\
$0.15$
\end{tabular}& 
\begin{tabular}{c}
$4.24$\\ 
$5.23$
\end{tabular}\\
\hline

93 & 
\begin{tabular}{c}
$1$\\
$2$
\end{tabular}&
\begin{tabular}{c}
$355$\\ 
$376$
\end{tabular}& 
\begin{tabular}{c}
$16.1$, $15.9$\\
$18.1$, $17.5$
\end{tabular}& 
\begin{tabular}{c}
$50.3$\\
$50.9$
\end{tabular}& 
\begin{tabular}{c}
$0.15$\\
$0.14$
\end{tabular}& 
\begin{tabular}{c}
$4.82$\\ 
$5.48$
\end{tabular}\\
\hline

105 & 
\begin{tabular}{c}
$1$\\
$2$
\end{tabular}&
\begin{tabular}{c}
$361$\\ 
$384$
\end{tabular}& 
\begin{tabular}{c}
$16.4$, $16.1$\\
$15.4$, $18.5$
\end{tabular}&
\begin{tabular}{c}
$50.5$\\
$45.6$
\end{tabular}& 
\begin{tabular}{c}
$0.21$\\
$0.08$
\end{tabular}& 
\begin{tabular}{c}
$5.90$\\ 
$4.30$
\end{tabular}\\
\hline

\end{tabular*}}
\caption{\label{tab:parameters} Mode characteristics for the $6$ selected nanoantennas.\\
Each of the characterized nanoantennas is treated as bimodal, with the modes corresponding to uncoupled Lorentzian resonances. The table provides fit parameters: central frequencies $\omega_\mathrm{j}$, widths of the scattering $\Gamma^\mathrm{scat}_\mathrm{j}$ and absorption $\Gamma^\mathrm{abs}_\mathrm{j}$ resonances, the resulting mode efficiencies $\eta_\mathrm{j}$, coupling constants $\kappa_\mathrm{j}$ to a quantum emitter located at the position indicated in Fig.~\ref{fig:nanoantenna2}(a), and Purcell enhancement factors $F$.}
\end{table}

\section{Emitted light} \label{sec:quantum}
Following Refs.~\onlinecite{Truegler2008} and \onlinecite{Waks2010}, we treat the nanoantenna resonances as quantum-mechanical modes, coupled to the quantum emitter with a transition frequency $\omega_\mathrm{qe}$, that belongs to the frequency range in which the nanoantenna modes overlap. The emitter is additionally subject to the free-space spontaneous emission and dephasing. Due to an incoherent excitation of the quantum emitter, through the pump $P$, energy is provided to the system in a quantized manner. This can be realized, e.g., through standard optical techniques \cite{Demtroder}. An excited emitter likely decays with a photon emission into either one of the two nanoantenna modes. The properties of the modes given in Table \ref{tab:parameters}, enter the quantum-mechanical Lindblad equation responsible for the dynamics and the stationary state of the coupled system. Such quantum description is discussed in detail in Appendix C,\comment{Ref.~\onlinecite Straubel2015} in which the stationary solution is sought.

\begin{figure}[h!]
\begin{centering}
\includegraphics[width=8.6cm,keepaspectratio]{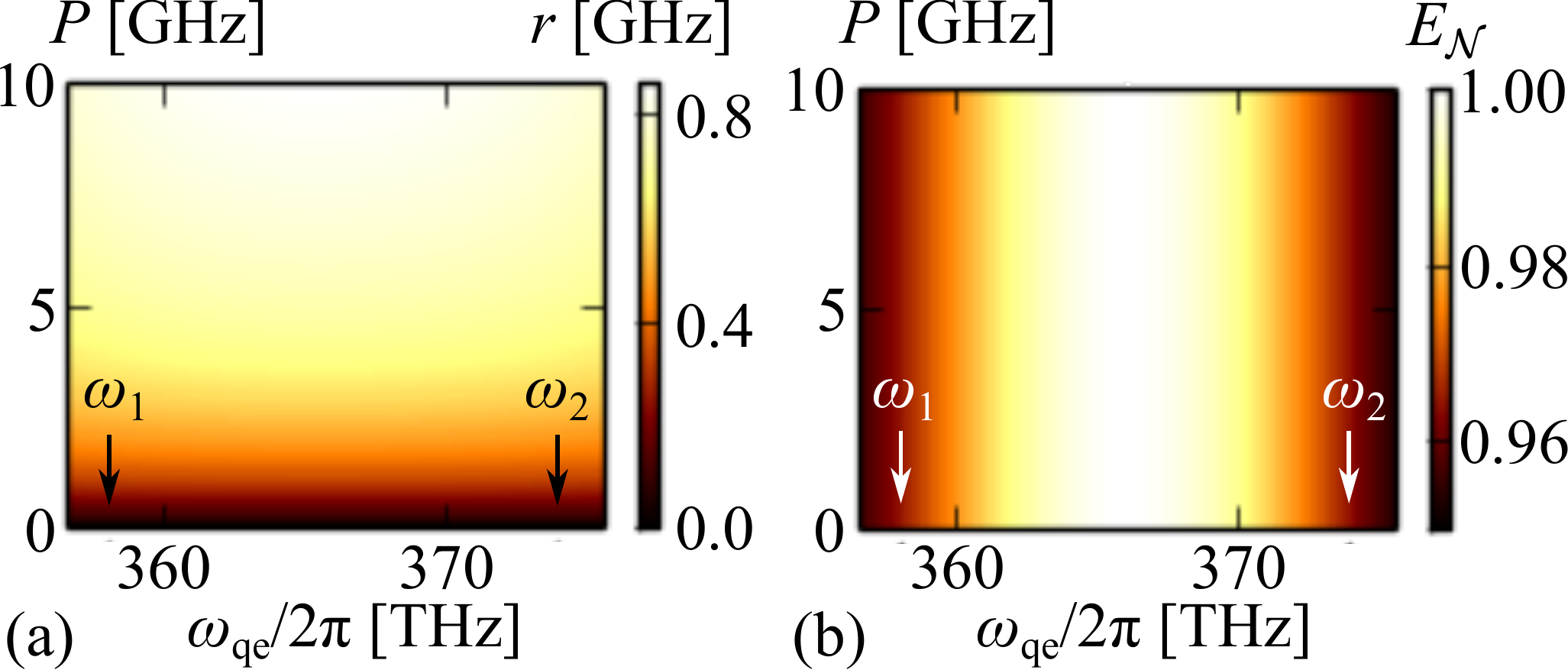}
\par
\end{centering}
\caption{\label{fig:results1} (a) Total photon emission rate $r$ and (b) logarithmic negativity $E_\mathcal{N}\left({\rho}_\mathrm{ff}\right)$, depending on $\omega_\mathrm{qe}$ and pump $P$. Results in (a) and (b) correspond to nanoantennas with $\Delta L=0$.}
\end{figure}

As a result, we find that realistic pump rates lead to photon emission rates $r=\sum_\mathrm{j} \Gamma^\mathrm{scat}_\mathrm{j}\langle a_\mathrm{j}^\dagger a_\mathrm{j}\rangle$ in the order of a GHz, i.e. comparable to the free-space spontaneous emission rate. Please note, however, that due to the nanoantenna, the directivity of the emission is related to the properties of the excited modes. Due to a potentially high directivity of nanoantennas \cite{Krasnok2014}, 
the detection efficiency could be significantly improved with respect to the free-space case. In Fig.~\ref{fig:results1}(a) we plot the rate $r$ for the set of parameters corresponding to the nanoantenna of $\Delta L=0$, and as a function of quantities that can be modified once a nanoantenna is fabricated: the pump $P$ and the transition frequency of the quantum emitter $\omega_\mathrm{qe}$. For this plot, the quantum emitter's dipole moment is fixed parallel to the $x$ axis, exciting both modes of the nanoantenna simultaneously. The results show a quite intuitive growth with the pump, which is however fairly independent on the quantum emitter transition frequency $\omega_\mathrm{qe}$. This is due to the broad character of the nanoantenna resonances. 

The corresponding level of entanglement can be expressed in terms of logarithmic negativity \cite{Vidal2002,Plenio2005} of the state of the bimodal light emitted into the far field $\rho_{\mathrm{ff}}$ (for a definition please see Appendix C). The entanglement measure takes the form
\begin{equation}
E_\mathcal{N}\left(\rho_{\mathrm{ff}}\right) = \mathrm{log} ||\rho_{\mathrm{ff}}^{\mathrm{T}_2}||,
\end{equation}
where $\mathrm{T}_2$ stands for partial transpose in mode $2$, and $|| ^\mathbf{.} ||$ represents the trace norm. This entanglement measure is applicable for systems of arbitrary dimensions, and therefore suitable in the case investigated here.

As a result, we find that once a photon is sent into the far-field, a high degree of entanglement in the emission modes is present. The result hardly depends on the applied pump, except for extremely weak pumps for which practically no photons are generated. The logarithmic negativity gently drops as the transition frequency of the quantum emitter shifts away from its optimal value, roughly about the mean frequency of the two modes [Fig.~\ref{fig:results1}(b)] (the exact value $\omega_\mathrm{qe}^\mathrm{opt}$ of the optimal $\omega_\mathrm{qe}$ is a function of the whole set of nanoantena parameters from Table~\ref{tab:parameters}, and may be slightly shifted from $\left(\omega_1+\omega_2\right)/2$ if the modes are not identical). Such weak dependence on $\omega_\mathrm{qe}$ indicates that there is no requirement to strictly match the resonances of the quantum emitter and the nanoantenna.

\begin{figure}[h!]
\begin{centering}
\includegraphics[width=8.6cm,keepaspectratio]{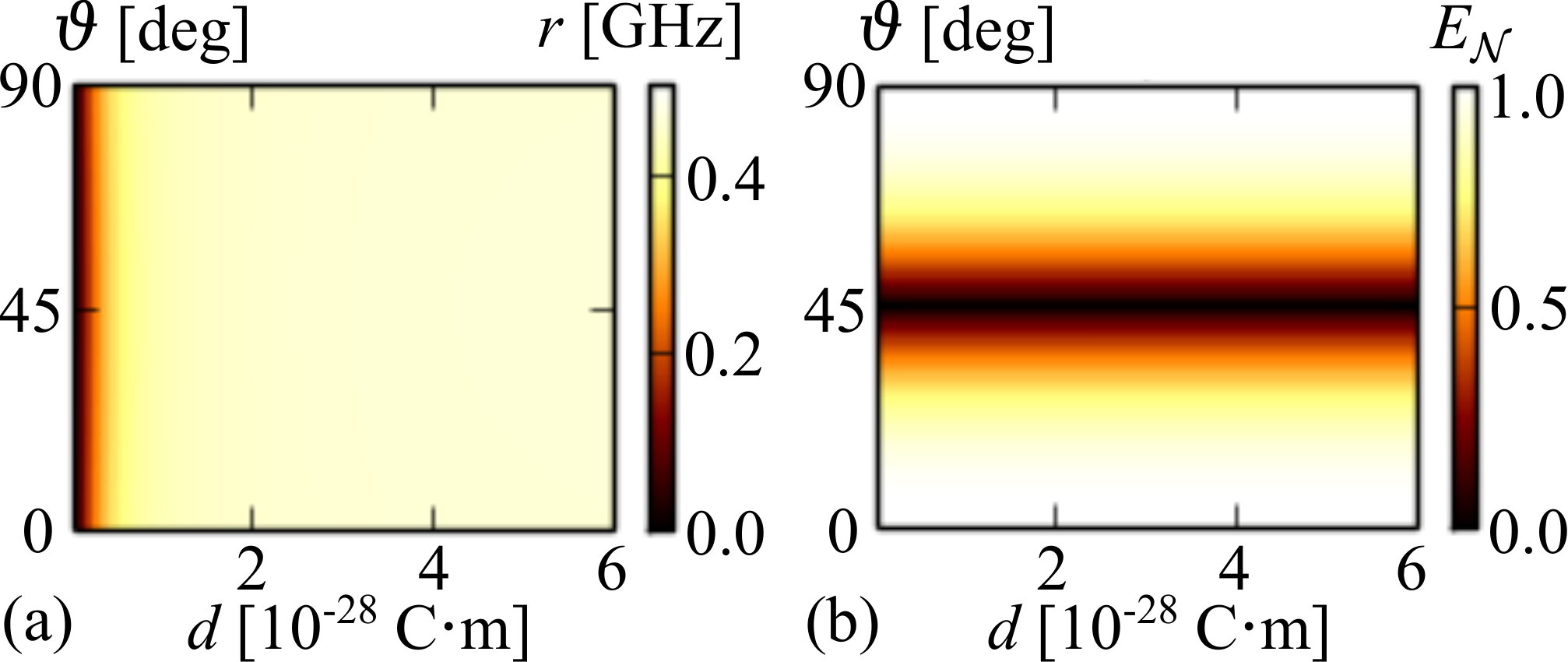}
\par
\end{centering}
\caption{\label{fig:results2} (a) Total photon emission rate $r$ and (b) logarithmic negativity $E_\mathcal{N}\left({\rho}_\mathrm{ff}\right)$ as functions of the dipole moment magnitude $d$ and orientation $\vartheta$, for $P=1$~GHz, and $\omega_\mathrm{qe}^\mathrm{opt}$. Results in (a) and (b) correspond to nanoantennas with $\Delta L=0$.}
\end{figure}

The dependence of the emission rate $r$ on the transition dipole moment $d$ of the quantum emitter [Fig.~\ref{fig:results2}(a)] can be intuitively explained as follows: In plasmonic cavities, the on-resonance nanoantenna-enhanced emission into the $\mathrm{j}$th mode grows with the square of the corresponding coupling constant $r_\mathrm{j} \sim |\kappa_\mathrm{j}|^2$. In the dipole approximation that is here applied, the latter is simply proportional to the dipole moment magnitude $d$ for each of the two modes. Therefore, we would expect a quadratic growth of $r$ with $d$. The saturation observed for larger dipole moments results from a finite value of the pump $P$ and finite efficiencies $\eta_\mathrm{j}$: the emission rate $r$ is limited by the rate $P$ at which the energy is provided to the system. The orientation $\vartheta$ of the dipole moment determines how this energy is eventually distributed among the modes, but due to their balanced emission properties, the total emission-rate dependence on $\vartheta$ is weak.

The degree of entanglement is robust against such rescaling of $d$ [Fig.~\ref{fig:results2}(b)]: Again, $\kappa_\mathrm{j}\sim d$, and the balance in properties of both modes leads to the high degree of entanglement. This balance occurs, naturally, only for selected orientations of the quantum emitter.  

\begin{figure}[h!]
\begin{centering}
\includegraphics[width=8.6cm,keepaspectratio]{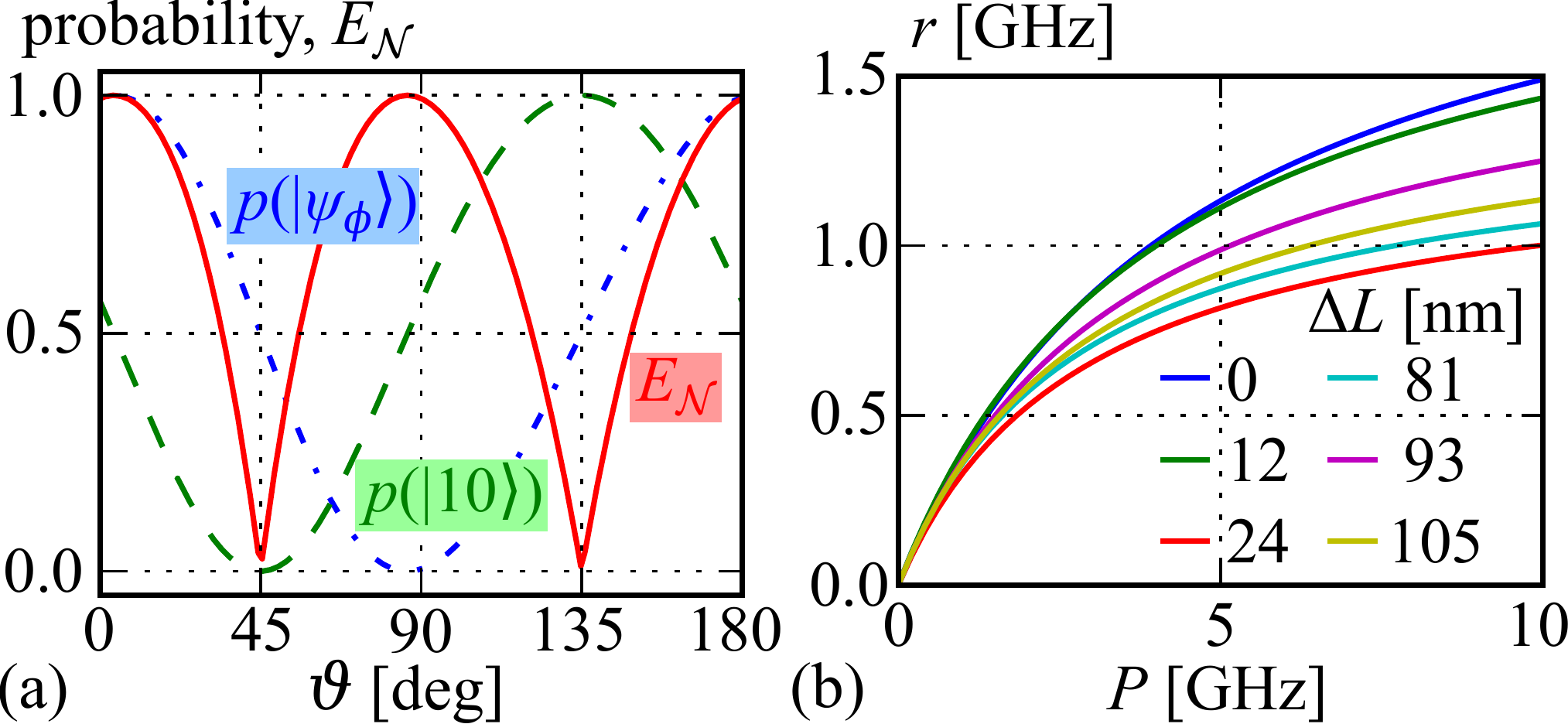}
\par
\end{centering}
\caption{\label{fig:results3} (a) $E_\mathcal{N}\left({\rho}_\mathrm{ff}\right)$ (red solid line), and probabilities of the Bell-like entangled state $|\Psi_\phi\rangle$ (blue dot-dashed line ) and the separable state $|10\rangle$ (green dashed line), as function of $\vartheta$ for nanoantennas with $\Delta L=0$. (b) Emission rate for the representative nanoantennas, depending on $P$ and for $\omega_\mathrm{qe}^\mathrm{opt}$.}
\end{figure}

In the next calculation, we confirm that the high degrees of entanglement originate from the emission of balanced superpositions $|\psi_\phi\rangle$, whose occurrence is very likely due to the similarity of the two nanoantenna modes. For Fig.~\ref{fig:results3}(a) we have chosen $\omega_\mathrm{qe}^\mathrm{opt}$ and a fairly low pump $P=0.1$ GHz. The figure shows the logarithmic negativity depending on the quantum emitter orientation $\alpha$. At the orientation parametrized by $\vartheta=45^{\circ}~\left(135^{\circ}\right)$, only mode 2 (1) is excited, leading to an emission of the separable state $|01\rangle$ ($|10\rangle$). The probability of the latter is plotted in the same figure with the green dashed line. Contrary, for dipole orientations parallel to one of the rods, one of the two orthogonal superposition states $|\psi_\phi\rangle$, $|\psi_{\phi+\pi/2}\rangle$ is excited (blue dot-dashed line corresponds to the probability $p\left( |\psi_\phi\rangle \right)$, the complementary one $p\left( |\psi_{\phi+\pi/2}\rangle \right)$ is phase-shifted by $\pi/2$ and not shown). 

Finally, we  compare the six nanoantennas, given in Table~\ref{tab:parameters}, in terms of the corresponding rate and degree of entanglement of the emitted light [Fig.~\ref{fig:results3}(b)]. For calculations we have again fixed the quantum emitter transition frequency, and for simplicity we have set $\omega_\mathrm{qe}^\mathrm{opt}$. As before, the results turn out to be very stable: for the worst (i.e. the lowest in brightness) nanoantenna of the set, we have obtained the emission rate reduced by about $30\%$ with respect to the best, symmetric one. That 'worst' nanoantenna is the one that corresponds to $\Delta L=24$ nm, whose spectrum is located furthest from the center of the resonance anticrossing in Fig.~\ref{fig:nanoantenna3}(a). While the rate of emission is a function of nanoantenna parameters, the logarithmic negativity hardly depends on these parameters in the range of pumps considered here. (To see a noticable difference we had to increase pump rates to unphysically large values, comparable to the quantum emitter's transition frequency.) This suggests robustness of the scheme proposed in this work, against possible imperfections.

We emphasize, that the state produced in our scenario corresponds to entanglement of two light modes in the degree of freedom related to the occupation number, rather than a more intuitive form of entanglement between two particles \cite{Horodecki2009}. This form of quantum correlations has been discussed in numerous proposals of Bell tests \cite{Tan1991} and other experiments devoted to proving single photon nonlocality (please see Ref.~\onlinecite{Hardy1994} and the discussion in the following comments). Another set of works discussed means to prepare \cite{Plenio2002}, process \cite{Mompart2003,Slowik2012}, and detect such states through single-photon interference experiments \cite{Hessmo2004}. Please compare also to the notion of time-bin qubits \cite{Donohue2013}. In Ref.~\onlinecite{Hessmo2004}, the single-photon nonlocality was finally demonstrated. 

\section*{Conclusions\label{sec:conclusions}}
Our contribution provides a detailed scheme to entangle two light modes in their degree of freedom related to the occupation number. This was possibly with a family of bimodal plasmonic nanoantennas made of two perpendicular nanorods. Due to the dominant value of nanoantenna scattering and absorption rates, a stable and bright emission is achieved in this scheme. Each time a photon is produced, the related degree of entanglement is high and robust: the scenario does not require high accuracy in the spectral properties of the selected quantum emitter, and is also rather insensitive to its additional decoherence, typically much smaller than the scattering rate of the investigated nanocavity. The results take into account the imperfect and unequal efficiencies of both modes and are stable in the investigated range of nanorod lengths, which suggests general robustness with respect to the nanoantenna geometry. The only requirement for the conceivable experiment seems to be a fairly large degree of control over the orientation of the quantum emitter's dipole moment, since for a significant degree of entanglement, specific orientations are necessary.

\section*{Appendix A: Simulation of spectra}
In this Appendix, a method to obtain the scattering spectra demonstrated in Fig.~\ref{fig:nanoantenna2}(b) and Fig.~\ref{fig:nanoantenna3}(a,b), is explained. In a similar way, absorption spectra can be calculated.

The task is to find the response of the nanoantenna to an excitation by a classical electric dipole, that represents an electric-dipole transition of a quantum emitter (for a similar treatment, please see Refs.~\onlinecite{Muskens2007} and \onlinecite{Taminiau2008}). The source is positioned at the vertex of the symmetry axes of the two nanorods, at the point indicated in Fig.~\ref{fig:nanoantenna2}(a) with the blue dot. As can be seen from Fig.~\ref{fig:nanoantenna2}(b) and Fig.~\ref{fig:nanoantenna3}(a,b), the response strongly depends on the orientation of the source, which therefore constitutes a knob to tune in and out of the individual mode resonances or address both modes simultaneously. Please note that the nanometric accuracy in the quantum-emitter positioning required for our scheme is feasible with several current experimental techniques \cite{Schietinger2009,Eizner2015,Chekini2015}.

Once the geometry of the nanoantenna and the source is fixed, we find the electromagnetic fields surrounding and within the nanoantenna by simulations of the full-vectorial Maxwell's equations in the frequency domain. We exploit here the commercially-available COMSOL Multiphysics simulation platform. The dispersive properties are included through the experimental data from Ref.~\onlinecite{Palik}.

\begin{figure}[t!]
\begin{centering}
\includegraphics[width=8.6cm,keepaspectratio]{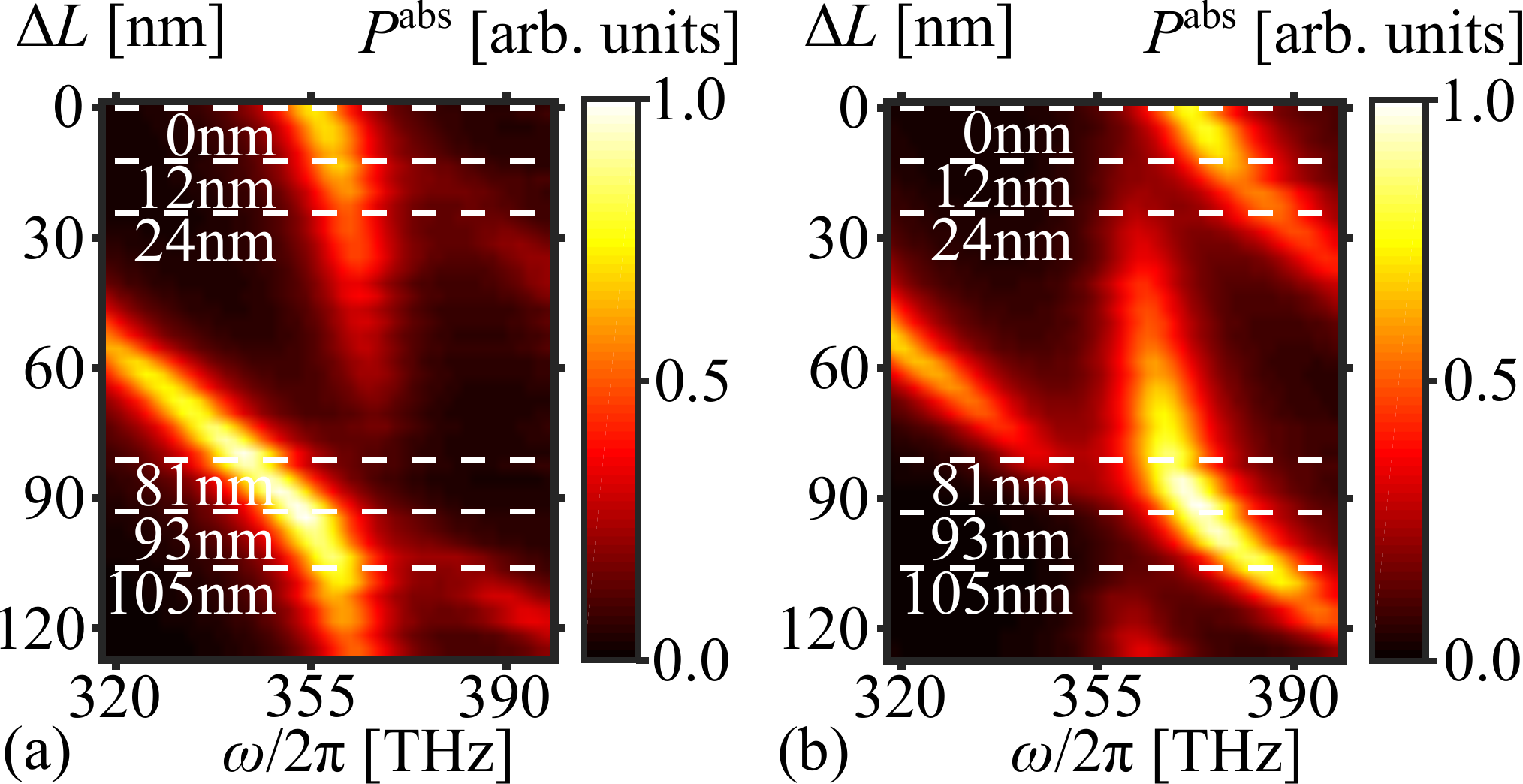}
\par
\end{centering}
\caption{\label{fig:AppA} 
(a) Absorption spectra $P^\mathrm{abs}(\omega)$ for electric dipolar sources oriented at $\vartheta=45^{\circ}$ and varying $\Delta L$. 
(b) As in (a), but with $\vartheta=135^{\circ}$.}
\end{figure}

Such field distribution is obtained twice: with and without the nanoantenna, where the difference corresponds to the fields scattered by the nanoantenna $\mathbf{E}^{\mathrm{scat}}(\mathbf{r},\omega)$ and $\mathbf{H}^{\mathrm{scat}}(\mathbf{r},\omega)$. Additionally, the same calculation provides the induced current distribution $\mathbf{j}^{\mathrm{ind}}(\mathbf{r},\omega)$. Please note that as long as the spatial dimensioning of the antenna design, i.e. size of the nanorods and distance to the quantum emitter, is well above the single nanometer regime, the hybrid system can be described by fully classical scattering simulations. Consequently we can employ the time averaged Poynting vector to formulate the conservation of energy and model the energy flux \cite{Giannini2011,Knight2009}.
The scattered and absorbed powers are obtained through Poynting's theorem:
\begin{eqnarray}
 P^\mathrm{scat}(\omega) &=& \oint_A \mathbf{S}^{\mathrm{scat}}(\mathbf{r},\omega)\cdot d\mathbf{A} \label{eq:P_scat}\\ 
 P^\mathrm{abs}(\omega)  &=& \int_V \mathbf{j}^{\mathrm{ind}}(\mathbf{r},\omega)\cdot \mathbf{E}^{\mathrm{inc}}(\mathbf{r},\omega)dV, 
\end{eqnarray}
where $\mathbf{S}^{\mathrm{scat}}(\mathbf{r},\omega)$ is the Poynting vector, integrated over a closed surface $\mathbf{A}$, and the volume $V$ denotes the bulk of the nanoantenna. The absorption spectra can be found in Fig.~\ref{fig:AppA}. Please note that the location of resonances is in agreement with the complementary scattering spectra from Fig.~\ref{fig:nanoantenna3}(a,b) from the main text.

We would like to draw the Reader's attention to the point that the two resonances discussed here are clearly not the fundamental modes of the individual nanorods,
but rather hybridized higher-order resonances, since the orientations of the dipole moment corresponding to the individual excitation of the modes do not coincide with the orientation of the dipole moment for a fundamental dipole excitation of a single nanorod, i.e. parallel to the long axis of the nanorod. A hybridization of the resonant modes of the two nanorods is expected, since the two rods are placed in close vicinity of each other. But since the only requirements for the two modes involved in the proposed entanglement generation scheme are their excitability and a spectral mode overlap enclosing the desired transition frequency of the quantum emitter, further characteristics of the modes constitute additional degrees of freedom with respect to potential applications. Similar results might be achievable with fundamental modes, however, this would typically require nanoantennas smaller in size, which might pose a greater technological challenge. Therefore, we have not restricted our calculations to the fundamental, dipolar modes of the nanorods.

The positions of the resonances and their overlap can be further tuned by a modification of other geometrical properties of the nanoantenna, such as the thickness of the rods, their size, shape, distance or embedding.

\section*{Appendix B: Nanoantenna characteristics}
Here, the method used to extract the nanoantenna characteristic parameters that later enter the quantum Lindblad equation, is discussed.

We base our general approach on a fit of the simulated nanoantenna scattering and absorption spectra with either Lorentzian or Fano line shapes\cite{Gallinet2011}. This is in accordance with experimental results in Ref.~\onlinecite{Verellen2014}, which have shown a presence of both symmetric Lorentzian profiles corresponding to isolated, and asymmetric Fano profiles corresponding to coupled modes in the spectra of golden nanorods.

\begin{figure}[h!]
\begin{centering}
\includegraphics[width=8.6cm,keepaspectratio]{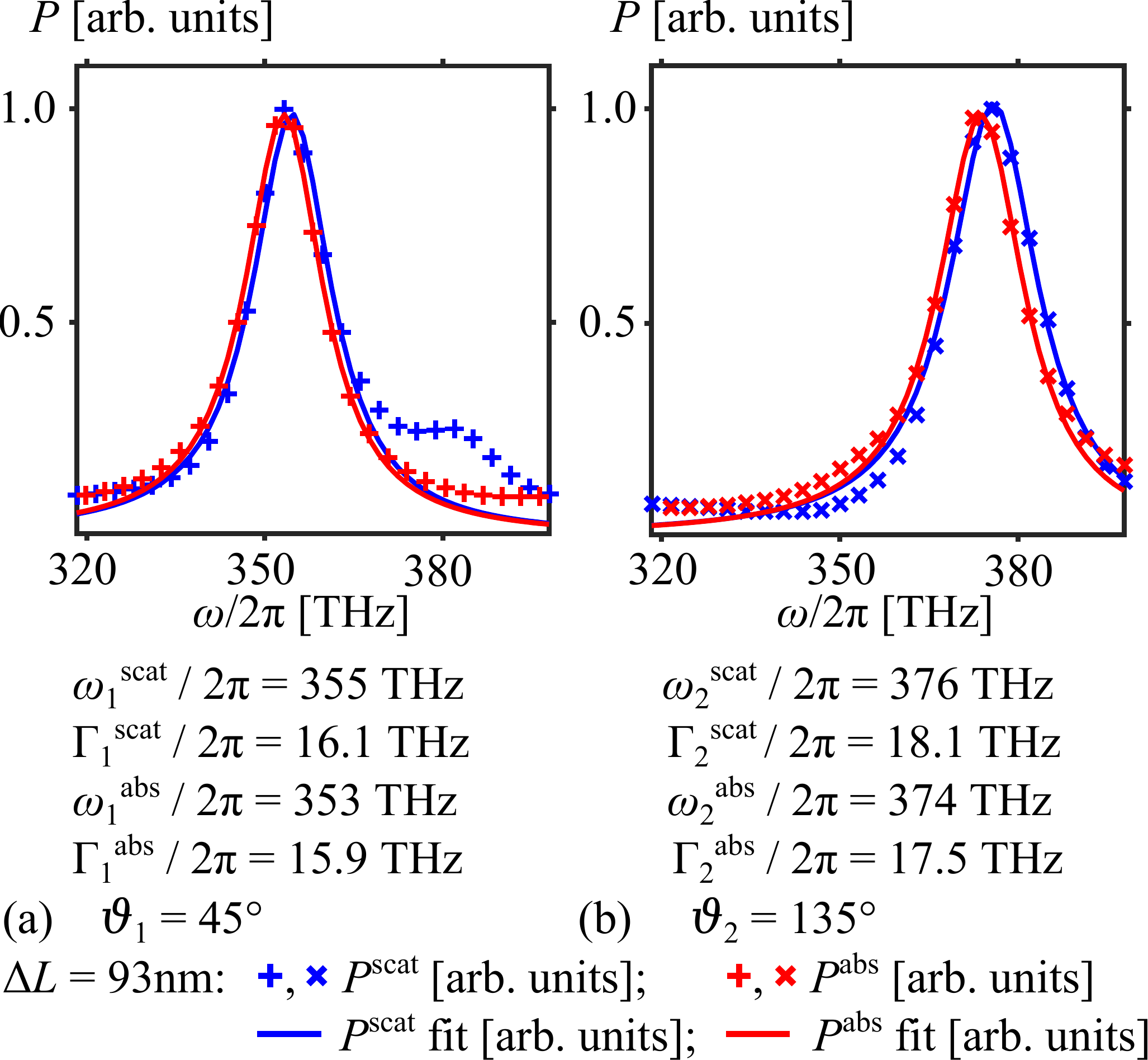}
\par
\end{centering}
\caption{\label{fig:AppB}
Spectra of scattered and absorbed electromagnetic power for individual mode resonance due to dipole excitation according to (a) $\vartheta=45^{\circ}$ and (b) $\vartheta=135^{\circ}$ for a nanoantenna of $\Delta L=93$nm. Additionally shown are Lorentzian fits of the line shapes and the characteristic parameters derived from these fits. Please note that the amplitudes have been scaled to $1$ for convenience, since thereby neither linewidth nor central frequency are affected.}
\end{figure}

In agreement with their quantum-mechanical representation as individual bosonic modes, we find that two uncoupled Lorentzians accurately represent the simulated spectra. Each Lorentzian profile $\mathcal{A}_\mathrm{j}\left(\omega\right)$ corresponds well to the case of a single driven, damped harmonic oscillator, decoupled from other resonances\cite{Joe2006}:
\begin{equation}
\partial^{2}_{t} x_\mathrm{j} + \Gamma_\mathrm{j}^\mathrm{k}\partial_{t} x_\mathrm{j} + \left.\omega_\mathrm{j}^\mathrm{k}\right.^{2}x_\mathrm{j} = \mathcal{E}^{\mathrm{k}}_\mathrm{j}\mbox{e}^{-i\omega t},  
\end{equation}
with \mbox{$x_\mathrm{j}\left(t\right) = \mathcal{A}^{\mathrm{k}}_\mathrm{j}\mbox{e}^{-i\omega t}$} and
\begin{equation}
\mathcal{A}^{\mathrm{k}}_\mathrm{j}\left(\omega\right) = \frac{\mathcal{E}^{\mathrm{k}}_\mathrm{j}}{\left(\omega_\mathrm{j}^\mathrm{k}\right)^{2} - \omega^{2} + i\Gamma_\mathrm{j}^\mathrm{k}\omega}.
\end{equation}
Here $\mathrm{j}$ is a numerical subscript denoting different resonances, while the superscript $\mathrm{k}$ differentiates between resonances in the scattered or absorbed fields $\mathrm{k} \in \lbrace \mathrm{scat},\mathrm{abs} \rbrace$. Such a fit directly provides the central frequency of the mode $\omega_\mathrm{j}$, as well as the scattering and absorption rates $\Gamma_\mathrm{j}^{\mathrm{k}}$. In the main text we have dropped the superscript $\mathrm{k}$ in the frequencies $\omega_\mathrm{j}$, since the resulting values coincide for the scattering and absorption cases with the accuracy of the order of $1\%$ (Fig.~\ref{fig:AppB}). 

Please note that the difference between the Lorentzian fit and the scattered power spectrum far detuned from the resonance frequency originates from contributions of other much further detuned modes to the scattered field. But these additional contributions are negligible on resonance with respect to the ascertainment of central frequency and linewidth. Furthermore, the single mode resonance orientations have been chosen to characterize the two modes with maximal accuracy. But the ability to individually excite each mode is no prerequisite with respect to the entanglement generation. 

For the calculation of the quantum state of light emitted from the setup, it is yet crucial to find the coupling strengths $\kappa_\mathrm{j}$ to the adjacent emitter. We obtain them basing on a comparison of the cavity-QED and classical expressions for the radiation enhancement of a source (so-called Purcell factors), which read in the multi-modal case:
\begin{eqnarray}
 F_\mathrm{cQED}  &=& 1+\sum_{\mathrm{j}} \frac{\eta_\mathrm{j}\kappa_\mathrm{j}^2\Gamma_\mathrm{j}}{[(\Gamma_\mathrm{j}/2)^2+(\omega_\mathrm{j}-\omega_\mathrm{qe})^2]\gamma_\mathrm{fs}},\\
 F_\mathrm{class} &=& 1+ \sum_\mathrm{j} f_\mathrm{j},
\end{eqnarray}
where the Weisskopf-Wigner free-space spontaneous emission rate of the quantum emitter is given by $\gamma_\mathrm{fs} = \frac{\omega_\mathrm{qe}^3\sqrt{\varepsilon}|d|^2}{3\pi\varepsilon_0\hbar c^3}$\cite{Scully}. Please note that we examine the influence of the dipole moment in Fig.~\ref{fig:results2}(a,b) of the main text. Otherwise, the value is fixed at $d=6\times10^{-29}$ C$\cdot$m. For each mode, the partial Purcell contributions $f_\mathrm{j}=\frac{P_\mathrm{scat,j}}{P_0}$ are defined as the ratio of the power scattered in the presence of the nanoantenna given by Eq.~(\ref{eq:P_scat}), and a corresponding power calculated in the absence of the nanoantenna. To find the coupling constants, we compare directly the on-resonance enhancement expressions for each mode:
\begin{equation}
 \kappa_\mathrm{j} = \sqrt{ \frac{f_\mathrm{j}\Gamma_\mathrm{j}\gamma_\mathrm{fs}}{4\eta_\mathrm{j}}}.
\end{equation}
Please note that such coupling constant is proportional to the dipole moment of the source through the Weisskopf-Wigner rate $\gamma_\mathrm{fs}\sim |d|^2$.

\section*{Appendix C: Quantum description}
Below we provide a step-by-step detailed description of the calculation of the quantum state of the emitted light, which directly allows one to obtain the emission rate and the degree of entanglement.

To account for a generation of light entangled in the number of photons in the two modes, we employ a quantized formulation, with the modes represented by annihilation operators $a_\mathrm{j}$, with $\mathrm{j}=1,2$, whose eigenfrequencies read $\omega_\mathrm{j}$. As follows from the simulation of the nanoantenna spectra, the two modes are independent from each other, i.e. are not directly coupled. They are however spectrally broad and interact to a quantum transition between the ground state $|g\rangle$ and an excited state $|e\rangle$ of a quantum emitter. According to the initial requirements, the transition frequency $\omega_\mathrm{qe}$ is enclosed in the frequency range in which the modes overlap. The coupling constant $\kappa_\mathrm{j}$ between each of the modes and the quantum emitter has been calculated in Appendix B.

The Hamiltonian of such a system is time-independent in the rotating wave approximation:
\begin{eqnarray}
\mathcal{H}/\hbar &=& \sum_{\mathrm{j}=1,2}\omega_\mathrm{j} a_\mathrm{j}^\dagger a_\mathrm{j} + \omega_\mathrm{qe}\sigma_+\sigma_- \\
&+& \sum_{\mathrm{j}=1,2}\kappa_\mathrm{j}\left(\sigma_+ a_\mathrm{j}+a_\mathrm{j}^\dagger \sigma_- \right), \nonumber
\end{eqnarray}
where $\sigma_-=|g\rangle \langle e|$ and $\sigma_+=\sigma_-^\dagger$, are the flip operators of the active transition of the quantum system. 

Additionally, we will incorporate time-irreversible processes, i.e.:
\begin{itemize}
 \item the incoherent pumping $\mathcal{L}_{\sigma_+}\left(P \right)$ of the quantum emitter,
 \item free-space spontaneous emission $\mathcal{L}_{\sigma_-}\left(\gamma_\mathrm{fs} \right)$ of the quantum emitter,
 \item dephasing $\mathcal{L}_{\sigma_+\sigma_-}\left(\gamma_\mathrm{d}\right)$ of the quantum emitter, which we have fixed at a rather large rate $\gamma_\mathrm{d}=100$ GHz, usual for quantum dots,
 \item decay of the nanoantenna modes $\mathcal{L}_{a_\mathrm{j}}\left(\Gamma_\mathrm{j}\right)$, with $\Gamma_\mathrm{j} = \Gamma^\mathrm{scat}_\mathrm{j}+\Gamma^\mathrm{abs}_\mathrm{j}$  through the far-field scattering and bulk absorption, according to the results of the calculations in Appendix B.
\end{itemize}
An implementation of the pump would be possible through the optical techniques \cite{Demtroder}, with the exciting beam strongly detuned from $\omega_\mathrm{qe}$. This would either prevent or allow one to spectrally filter a direct excitation of the nanoantenna, detrimental for the emission statistics. 

These mechanisms are accounted for via the Lindblad operators:
\begin{equation}
\mathcal{L}_c\left(\gamma \right)\rho (t) = \gamma\left\lbrace c \rho(t) c^\dagger - \frac{1}{2}\left[c^\dagger c \rho (t)+\rho (t) c^\dagger c \right]\right\rbrace,
\end{equation}
where $\rho(t)$ represents the density matrix of the combined system of nanoantenna modes and the quantum emitter, $c$ is the operator responsible for the particular incoherent channel, and $\gamma$ is the rate of the corresponding process. 

The state $\rho$ of the system is found from the stationary form of the Lindblad equation:
\begin{equation}
0 = -i/\hbar \left[\mathcal{H},\rho \right] + \mathcal{L}\rho,
\end{equation}
where $\mathcal{L}$ is a direct sum of all above-described Lindblad operators. We find the stationary state $\rho$, using the freely-available, open-source QuTiP2 toolbox in Python \cite{qutip2}. This directly provides the photonic emission rates $r$ discussed in the main text.

\begin{figure}[h!]
\begin{centering}
\includegraphics[width=8.6cm,keepaspectratio]{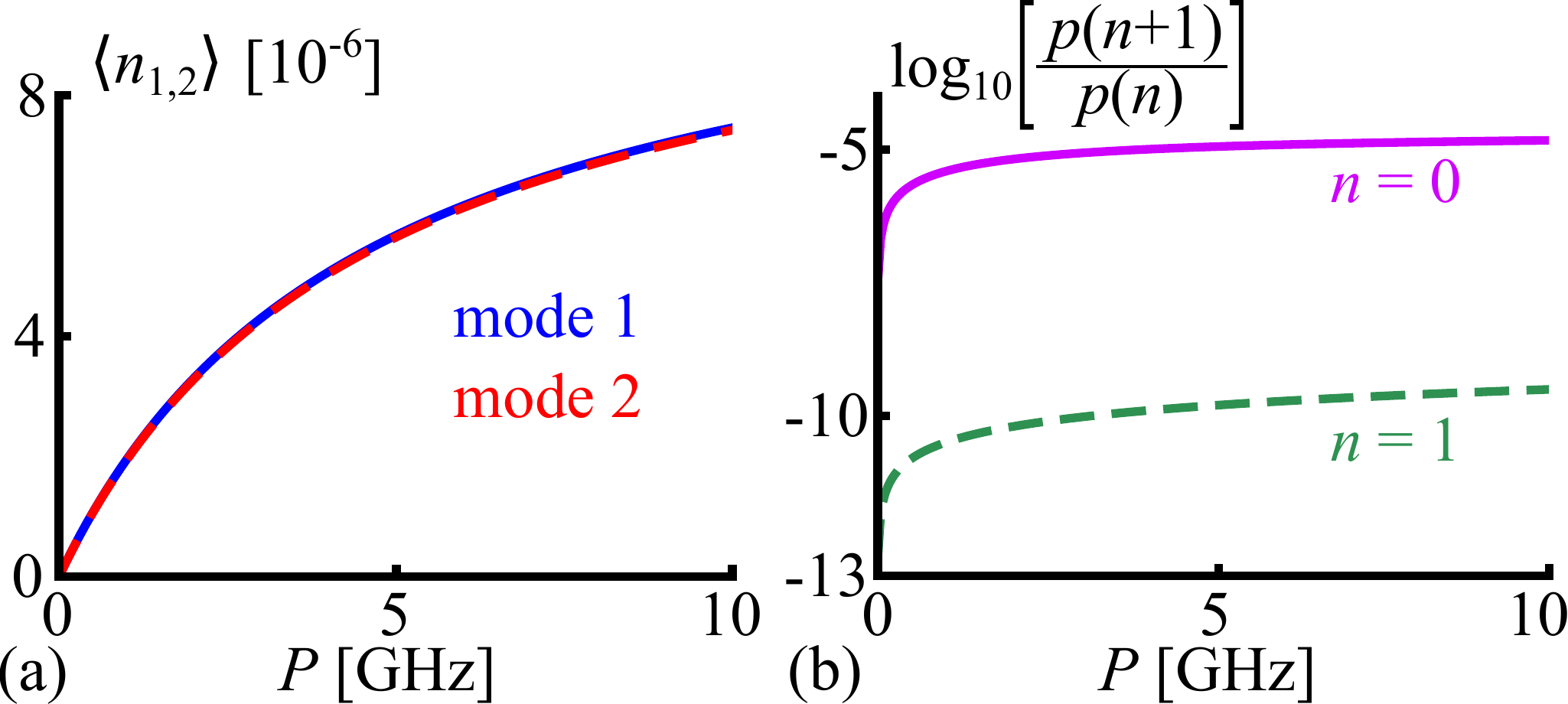}
\par
\end{centering}
\caption{\label{fig:AppC} (a) Stationary expectation values of the number of photons in modes $1$ (blue solid line) and $2$ (red dashed line), as functions of the pump rate $P$.
(b) Photon number ratios $\log_{10}\left[\frac{p(1)}{p(0)}\right]$ (purple solid line) and $\log_{10}\left[\frac{p(2)}{p(1)}\right]$ (green dashed line),
where $p(n)$ denotes that exactly $n$ photons in total are present in the system.
The plots correspond to the equal-arms nanoantenna ($\Delta L=0$), for $\omega_\mathrm{qe}=(\omega_1+\omega_2)/2$, and the dipole moment oriented along one of the nanorods.
}
\end{figure}

For realistic pumps in the GHz range, the typical mean photon numbers in the system are of the order of $10^{-5}$ [Fig.~\ref{fig:AppC}(a)]. The reason of such a small occupation number are the enormous scattering and absorption rates of the nanoantenna. and the probabilities of a total number of photons larger than $1$ are further $10$ orders of magnitude lower [Fig.~\ref{fig:AppC}(b)]. For this reason, it is sufficient to truncate the Hilbert space at photon numbers as low as $3$ for each mode. Larger spaces produce indistinguishable results. 

Since we focus on the state of light, we will limit the discussion to the density matrix of the two photonic modes, and trace out the degrees of freedom of the quantum emitter: $\rho_\mathrm{ph} = \mathrm{Tr}_\mathrm{qe}\left(\rho\right)$.

Small numbers of photons in the setup indicate that the vacuum state $|00\rangle$ is the dominant contribution to $\rho_\mathrm{ph}$. The overall values of any entanglement measure applied to the photonic state will naturally be very low. However, the vacuum state can hardly be properly detected or used for applications. For this reason, we focus on contributions, whose overall photon numbers are greater than zero, resulting in a detector click. We eliminate $|00\rangle$ from our considerations, and suitably normalize the resulting density matrix: 
\begin{equation}
\tilde{\rho} = \mathcal{N}\left(P_{\perp \mathrm{vac}} \rho_{\mathrm{ph}}P_{\perp \mathrm{vac}}\right),
\end{equation}
where $P_{\perp \mathrm{vac}} = \unit - |00\rangle \langle 00|$ is a projector on the non-vacuum subspace, and $\mathcal{N}\left(\rho\right)\equiv \frac{\rho}{\mathrm{Tr}\rho}$ assures a proper normalization. This operation implies that we only consider the states in which at least one photon is present.

It is left to include the imperfect efficiencies of the modes: some photons generated in the setup are absorbed rather than emitted into the far field. Since $\eta_\mathrm{j}$s are in general unbalanced (see Tab.~\ref{tab:parameters}), $\rho_\mathrm{ph}$ will differ from the state emitted into the far-field. In general, the transformation to the far-field density matrix $\rho_\mathrm{ph}$ can be derived using the input-output formalism, with a rather cumbersome result. In the investigated case, however, contributions from photon numbers higher that $1$ are suppressed by many orders of magnitude. Therefore, we can apply directly an intuitive transformation:
\begin{equation}
\rho_{\mathrm{ff}} = \mathcal{N}\left(T_\eta \tilde{\rho} T_\eta\right),
\end{equation}
where $T_\eta = \sqrt{\eta_1}|10\rangle \langle 10|+\sqrt{\eta_2}|01\rangle \langle 01|$. It simply multiplies a probability of a photon in the mode $\mathrm{j}$ by $\eta_\mathrm{j}$, and the corresponding coherence by $\sqrt{\eta_1\eta_2}$, while the trace operation keeps the resulting density matrix normalized.

The degree of entanglement of such far-field photonic state is investigated in detail in the main part of this contribution.

\section*{Funding Information}
The study was supported by a research fellowship within the project 
``Enhancing Educational Potential of Nicolaus Copernicus University in the Disciplines of Mathematical and Natural Sciences'' 
(project no. POKL.04.01.01-00-081/10.) and the Karlsruhe School of Optics and Photonics (KSOP). 
Partial support by the German Science Foundation (within SPP 1391 Ultrafast Nano-optics) is acknowledged.

\bibliography{bimodal_entangle}

\end{document}